\documentclass[preprint]{aastex}

\received{}
\revised{}
\accepted{}
\journalid{}
\articleid{}
\paperid{}
\ccc{}
\begin{document}

\newcommand{\etal}{{\em et~al.\,}}
\newcommand{\CL}{{RX J0848+4456}}
\newcommand\fdeg{\mbox{$.\!\!^{\circ }$}}

\title{\CL: Disentangling a Moderate Redshift Cluster\altaffilmark{1}}
\altaffiltext{1}{Based on observations with the NASA/ESA Hubble Space
Telescope, obtained at the Space Telescope Science Institute, which is
operated by the Association of Universities for Research in Astronomy,
Inc. under NASA contract No. NAS5-26555.}  

\author{B. P. Holden\altaffilmark{2}, S. A. Stanford\altaffilmark{2}}
\affil{Department of Physics, University of
California, Davis, CA 95616}
\email{bholden@igpp.ucllnl.org,adam@igpp.ucllnl.org}
\altaffiltext{2}{Participating Guest, Institute of Geophysics and
Planetary Physics, Lawrence Livermore National Laboratory}

\author{P. Rosati}
\affil{European Southern Observatory, Karl-Scharzschild-Strasse 2,
D-85748 Garching, Germany}
\email{prosati@eso.org}

\author{G. Squires}
\affil{California Institute of Technology, 105-24 Caltech, 1201 East
California Blvd, Pasadena, California 91125}
\email{gks@astro.caltech.edu}

\author{P. Tozzi}
\affil{Osservatorio Astronomico di Trieste, via G.B. Tiepolo 11,
I-34131, Trieste, Italy}
\email{tozzi@ts.astro.it}

\author{R. A. E. Fosbury}
\affil{Space Telescope-European Coordinating Facility, D-85748
Garching bei Munchen, Germany}
\email{rfosbury@eso.org}

\author{C. Papovich\altaffilmark{3}}
\affil{Dept. of Physics and Astronomy, The Johns Hopkins University, 
Baltimore, Maryland 21218}
\email{papovich@stsci.edu}
\altaffiltext{3}{Space Telescope Science Institute, The Space
Telescope Science Institute is operated by the AURA, Inc., under
National Aeronautics and Space Administration (NASA) Contract NAS
5-26555.}

\author{P. Eisenhardt}
\affil{Jet Propulsion Laboratory, California Institute of Technology,
MS 169-327, 4800 Oak Grove Drive, Pasadena, CA 91109}
\email{prme@kromos.jpl.nasa.gov}

\author{R. Elston}
\affil{Department of Astronomy, University of Florida, 211 Space
Sciences Bldg., Gainesville, FL 32611}
\email{elston@astro.ufl.edu}

\and 

\author{H. Spinrad}
\affil{Astronomy Department, University of California, Berkeley, CA 94720}
\email{spinrad@bigz.berkeley.edu}

\begin{abstract}

We present a multi-wavelength study of \CL, a cluster of galaxies
discovered through X-ray emission in the \texttt{ROSAT} Deep Cluster
Survey.  Our observations consist of WFPC2 imaging, optical spectra,
and X-ray data collected with the Chandra observatory.  We find that
\CL\ consists of an X-ray emitting cluster of galaxies at a redshift
of $z=0.570$ and a group at slightly lower redshift, $z=0.543$, with
little X-ray emission.  This lower redshift system, however, is a
gravitational lens, with the lensed galaxy an unusual AGN or
star-forming system at $z=3.356$.

The cluster has an X-ray temperature of kT = $3.6 \pm 0.4$ keV, a
bolometric luminosity of $1.0 \pm 0.3 \times 10^{44}$ erg s$^{-1}$ and
a velocity dispersion of $670 \pm 50\ {\rm km\ s^{-1}}$.  These values
all agree with the low redshift correlations for clusters of galaxies,
implying a relaxed system with the ICM in equilibrium with the dark
matter potential.  The lower redshift group of galaxies at $z=0.543$
has, at most \slantfrac{1}{5}\, more likely \slantfrac{1}{10}, of the
X-ray luminosity of \CL.  Despite being a gravitational lens, this is
a low mass system, with an X-ray temperature of kT =
$2.3^{+0.5}_{-0.4}$ keV and a velocity dispersion of only $430 \pm 20\
{\rm km\ s^{-1}}$.  Our observations show the importance of
detailed studies of clusters of galaxies when using them as probes of
cosmological mass functions.

\end{abstract}

\section{Introduction}

In order to understand the formation and evolution of clusters of
galaxies, not only are large samples needed, but the clusters in those
samples must be well understood.  Theoretical predictions for clusters
of galaxies usually present the space density of objects as a function
of mass, redshift, and cosmology.  Clusters are, however, found by
their X-ray luminosity, optical richness, or weak lensing signal.
Given a complete sample of clusters of galaxies found through one of
these methods, it is necessary to conduct a variety of follow-up
observations to understand the relation between the observed
quantities and the binding mass.  Only then can meaningful comparisons
with theoretical predictions be made.

We present here our follow up observations of \CL, a cluster
discovered in the \texttt{ROSAT} Deep Cluster Survey
\citep[RDCS]{rosati98} with a flux of $3.4 \times\, 10^{-14}$ erg
s$^{-1}$ cm$^{-2}$ (0.5--2.0 keV, observed) at 08$^{\rm h}$ 48$^{\rm
m}$ 46\fs 88 and $\delta$ = 44\arcdeg\ 56\arcmin\ 21\farcs 7 (J2000).
We have acquired X-ray data from Chandra, deep HST and ground based
imaging, and redshifts of cluster members from the Keck Observatory.
Each of these datasets, discussed in \S 2, sample different aspects of
the cluster's dark matter potential.  The X-ray data probes the hot
intra-cluster medium, the optical imaging is used for measuring the
projected cluster potential through gravitational lensing, and the
redshifts sample the potential along the line of sight.  We will
show our results in \S 3.  We have learned that any individual data
set would give us a skewed picture: all of the data sets need to be
examined to model this system as shown in \S 4.  We summarize our
results and discuss the ramifications this moderate redshift cluster
might have for other cluster surveys in \S 5.  We use $H_0 = 65\ {\rm
km\ s^{-1}\ Mpc^{-1}}$, $\Omega_{m} = 0.3$ and $\Omega_{\Lambda} =
0.7$ for our cosmological parameters.  At the redshift of interest,
this corresponds to approximately 7 kpc per arcsecond.

\section{Data}

\subsection{Optical Spectra}

For the first spectroscopic observations of \CL, done in November of
1998, one of us (GKS) selected galaxies from an I band image observed
with the Prime Focus camera the KPNO Mayall 4m.  Galaxies were
selected by eye from a 8\arcmin\ by 6\arcmin\ section, the size of the
Low Resolution Imaging Spectrograph \citep[LRIS]{oke95} field of view,
based on their size and apparent magnitude.  For the second set of
spectroscopic observations, taken in February of 2000, we supplemented our
original I band image with a R band image taken with LRIS at the Keck
observatory.  We used the R-I color to find the red sequence usually
found in clusters as a means of selecting candidates for that last
observing run.  The selected galaxy sample is not complete for any of
the three selection criteria.

We used LRIS on the Keck I telescope to observe spectra of 71 galaxies
with three different slit masks, between the two sets of observations,
using the 600 line mm$^{-1}$ grating blazed at 7500 \AA . This grating
produces $\sim$ 1.3 \AA\ per pixel resolution for our spectra. We
observed each mask multiple times with 1200 to 1800 second
observations, and made offsets across the dispersion direction to
dither the observations.  The total integration times for the three
masks ranged from 4800
seconds in good conditions to 7200 seconds in the worst conditions.
For this rest of the paper we will refer to these as the high
dispersion or high resolution spectra.

We reduced the two-dimensional data using a set of IRAF\footnote {The
Image Reduction and Analysis Facility (IRAF) software is provided by
the National Optical Astronomy Observatories (NOAO), which is operated
by the Association of Universities of Research in Astronomy for
Research in Astronomy, Inc., under contract to the National Science
Foundation.} scripts optimized for LRIS observations.  These scripts
greatly reduce the fringing evident in LRIS spectra at long
wavelengths when combined with dithered observations.  We extracted
one-dimensional spectra using the IRAF package APEXTRACT.

Using the RVSAO package \citep{kurtz98}, we measured redshifts of the
extracted spectra with the cross-correlation technique.  We used 
{\tt fabtemp97} \citep[which contains details on this template]{kurtz98} as
the cross-correlation template.  We successfully measured redshifts
for 62 galaxies in our sample. Of these 62, 34
had correlation values greater than two.  The remaining 28
redshifts were determined by visual examination and fitting absorption
or emission features.  Of those 28 spectra, 20 were pure
emission spectra while the remaining eight were fit with a combination
of emission and absorption features.  The average error on the
redshift estimates, regardless of whether or not they used the results
from RVSAO, is $\sigma_z = 2.7\times 10^{-4}$ or 81 km s$^{-1}$.

In addition to these targeted spectra, \CL\ falls within the Lynx
field of a field galaxy sample from the Spectroscopic, Photometric,
Infrared-Chosen Extragalactic Survey
\citep[SPICES]{eisenhardt_spices}.  Specifically, the Lynx portion is
a 5\arcmin\ by 5\arcmin\ field centered at $\alpha$ = 08$^{\rm h}$
48$^{\rm m}$ 44 \fs 8 and $\delta$ = 44\arcdeg\ 54\arcmin\ 25\farcs 9
(J2000).  Many cluster members have been observed as part of the
spectroscopic part of this study using LRIS to observe all $K<20$
galaxies with the 150 line mm$^{-1}$ grating, which yields a
resolution of 4.8 \AA\ pixel$^{-1}$.  These spectra, which we will
refer to as the low resolution sample, are useful for determining
cluster membership but not velocity dispersions.  The spectral data
for the SPICES survey was reduced using a similar set of IRAF scripts
as the 600 line mm$^{-1}$ grating data. As of the write of this paper,
we have 144 redshifts in the Lynx field of SPICES, though this
includes all spectra taken at the higher resolution that both fall
within the SPICES survey field and magnitude limit.

\subsection{HST Imaging}

Besides the ground based imaging, there are two sets of HST WFPC2
mosaics for \CL.  Both data sets used the F702W filter and had a total
of ten 1200 second exposures.  The northern set of HST images,
observed on April 20, 1999, are centered on $\alpha$ = 08$^{\rm h}$
48$^{\rm m}$ 49\fs 55 and $\delta$ = 44\arcdeg\ 57\arcmin\ 19\farcs 38
(J2000), while the southern set, observed on April 30, 2000, was
centered on $\alpha$ = 08$^{\rm h}$ 48$^{\rm m}$ 49\fs 44 and $\delta$
= 44\arcdeg\ 54\arcmin\ 38\farcs 81 (J2000).  The northern set had a
roll angle of 288\arcdeg\ while the southern set had a roll angle of
284\arcdeg.  The combination of the two sets of exposures samples the
core of the cluster while giving a good coverage of the less dense
outer regions.

We reduced the WFPC2 data using the tools provided by the
IRAF/STSDAS\footnote {Space Telescope Science Data
Analysis System (STSDAS) is distributed by the Space Telescope Science
Institute.}  standard pipeline software, including the DITHER and
DITHERII packages \citep{fruchter98}.  We fit and removed the
background sky level from each chip of the separate WFPC2 images.
Next, to identify bad pixels, we used the information in the image
data quality files, and we took advantage of the information from the
multiple dither positions to identify cosmic rays. We
cross--correlated each chip of the separate images with a common
reference position to derive dither offsets between each image.  Using
these offsets, we used the \texttt{drizzle} task to align the images,
and generated median images from these coaligned images.  We then
shifted the median images back to the original positions via the
\texttt{blot} task, and used the \texttt{driz\_cr} task to compare
them to the original exposures to identify cosmic rays and hot pixels.
Masking out the identified bad pixels and cosmic--rays, and using the
derived image offsets, we drizzled each of the four chips from the ten
separate exposures to produce a single final output image with a pixel
scale of $0\farcs046$~pix$^{-1}$, the scale of the PC chip in the
WFPC2 instrument.  In Figure \ref{cl2_xray_cont},
we show a subset of the data centered on the core of the cluster.

\subsection{X-ray Imaging and Spectra}

\CL\ was observed as part of a $\sim$190 ks Chandra ACIS-I observation
of the Lynx field.  The X-ray observations were broken up into two
parts, one of 62.1 ks (Obs ID 1708) taken on May 3, 2000 and one of
126.7 ks taken on May 4, 2000 (Obs ID 927).  Each observation was done
with the faint mode when ACIS was operating at a temperature of -120
C.

We processed the Level 1 data for each observation using the CIAO
v1.1.5 software. We cleaned the Level 1 event list for all events with
ASCA grades of 1, 5 and 7, filtered the data not in good time
intervals, removed bad offsets and removed bad columns.  We then
removed, on a chip by chip basis, 3.3 second time intervals when the
count-rate exceeded three standard-deviations above the average
count-rate, a rate of $\simeq$1.25 counts per second.  Once we had
created two clean event files and corresponding exposure maps, we
merged the two exposures into one event file for the whole pointing.
For both the Level 1 processing and the spectral analysis, we used the
ACIS calibration files that were available on September 15, 2000.

\section{Physical Measurements}

\subsection{Cluster Redshift and Velocity Dispersion}

We used the 600 line mm$^{-1}$ grating redshifts measured in \S 2.1 to
determine the redshift and velocity dispersion of \CL.  First, we used
the gap technique of \citet{katgert96} to define groups with similar
redshifts.  This technique simply groups galaxies where the difference
in the redshifts between group members is less than a cutoff.  If more
than a certain number of galaxies have spacings smaller than the
cutoff, that is declared a group along the line of sight.  No spatial
position information is used.  We chose the parameters of
\cite{adami00}, {\it i$.$e.\ }a gap of 1000 ${\rm km\ s^{-1}}$ in the
rest frame of the galaxies and a group size of five or more galaxies.
Using those parameters, we find two groupings, one with 17 galaxies
(with 12 absorption-line and five emission-line redshifts) and one with 11
(with nine absorption-line redshifts, one emission-line redshift and one
combination).  The remaining galaxies are spread uniformly over $0.151
\le z \le 1.173$.

\begin{figure}
\begin{center}
\includegraphics[height=6.0in]{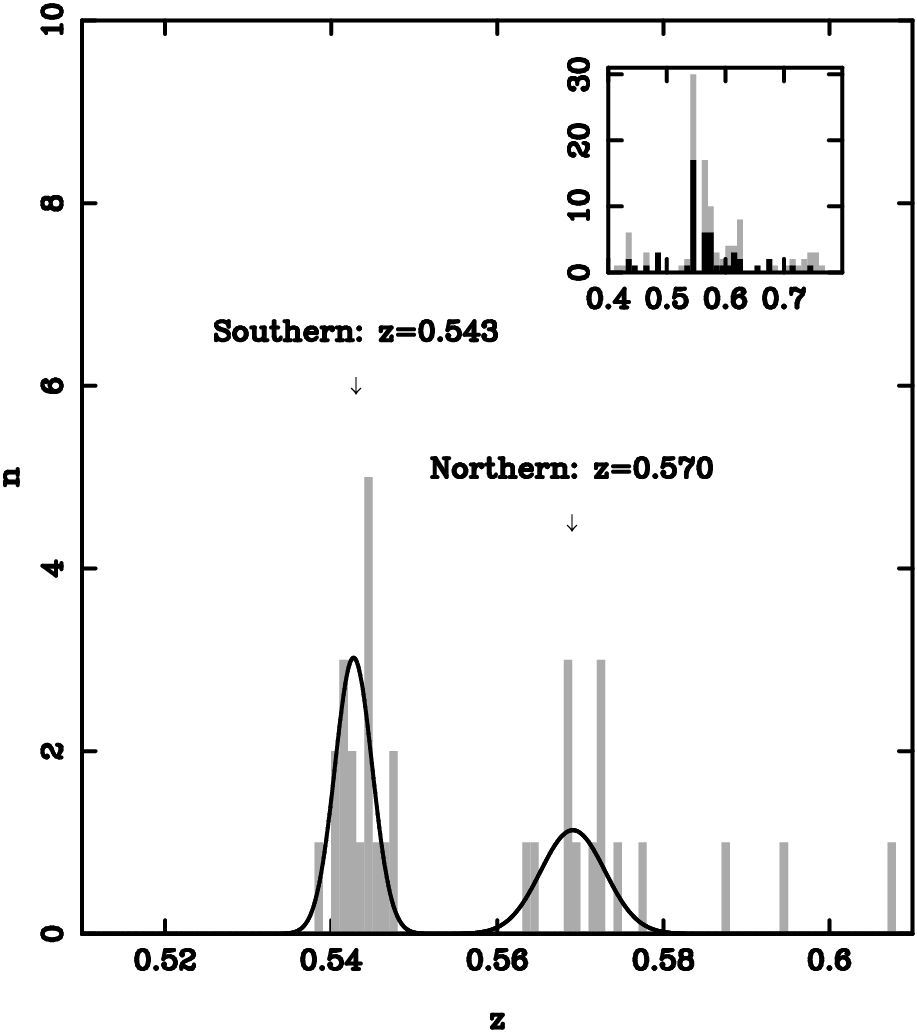}
\end{center}
\caption[cl2.fig2_pps.ps]{ Redshift distribution along the line of
sight towards RX J0848+4456 from the 600 line mm$^{-1}$ data plot with
bins of a width $\delta_z = 0.001$.  We mark the two groups found by
the gap algorithm discussed in \S 3.2, and plot Gaussians representing
the best fitting mean redshifts and velocity dispersions.  The inset
histogram, with bins of a width $\delta_z = 0.01$, shows both the
distribution of low (grey) and high (black) resolution spectra.
\label{zhistogram}}
\end{figure}

For each grouping, we used the biweight estimators of \citet{beers90}
to determine the cluster redshift and velocity dispersions.  We find
that the grouping of 17 galaxies is at a redshift of $z=0.543$ and the
grouping of 11 has a redshift of $z=0.570$.  To estimate the velocity
dispersions, we used the biweight scales from \citet{beers90}.  We
calculated a value of $430 \pm 20\ {\rm km\ s^{-1}}$ for the lower
redshift system and $670 \pm 50\ {\rm km\ s^{-1}}$ for the higher
redshift system.  We computed the errors on the velocity dispersions
using a jackknife estimate \citep{beers90}.  In Figure
\ref{zhistogram}, we plot the velocity distribution along the line of
sight with the two groupings marked.  Over the histogram, we plot two
Gaussians with a mean and variance specified by the biweight values
above.  As an inset, we plot both the high dispersion redshifts along
with the lower dispersion redshifts from SPICES with bins ten times
larger.  The high dispersion spectra are in bins colored black while
the low dispersion spectra are in grey bins.  There are a total of 32
spectra in the lower redshift grouping and and 25 in the higher
redshift system out of the total 110 redshifts plotted (including both
low and high resolution spectra).  The remaining 34 redshifts
available in this field are at redshifts outside the region displayed
in the inset.

The difference in redshift between these two systems is $\delta
z=0.0265$, which is significant.  This is a velocity difference of
$\sim$ 5200 km s$^{-1}$ in the frame of the two systems. Such a large
difference implies that they are not two parts of a merging system.

Determining the location on the sky of these two groupings is
problematic.  We selected galaxies based on imaging data centered to
the south of the cluster.  No galaxies in our spectroscopic sample lie
more than 1\farcm 1 north of the original RDCS centroid.
Therefore, when we calculate a centroid of the galaxies with redshifts
in each grouping we are biased towards values to the south.

The brightest member of the $z=0.570$ grouping lies 10\farcs 9 (77
kpc) to the east of the peak in the X-ray emission in Figure
\ref{cl2_xray_cont}.  At the peak is a slightly fainter galaxy that
has an extended envelope, typical of the bright, early-type galaxies
at the cores of clusters.  We deem it likely that the center of the
$z=0.570$ grouping lies near one of these two galaxies.  There is an
equivalent bright, early-type galaxy at $z=0.543$ at the center of a
fainter X-ray peak to the southeast in Figure \ref{cl2_xray_cont}.
This galaxy is likely at or near the center of the lower redshift
system.

\subsection{Chandra Results}

In Figure \ref{cl2_xray_cont}, we plot the X-ray contours over the HST
optical image.  In addition to the main peak in the X-ray emission,
near a bright galaxy to the southeast is a second peak in X-ray
emission.  This peak is much fainter than the first peak, however it
is near a galaxy at a redshift of $z=0.543$, while the galaxies at
the peak of the main X-ray emission are all at $z\sim0.57$.

We extracted all events in an aperture of 100\arcsec\ around the main
peak in the X-ray emission.  Using the extracted events, we measured
the centroid of the X-ray emission and found that to be $\alpha$ =
08$^{\rm h}$ 48$^{\rm m}$ 47\fs 83 and $\delta$ = 44\arcdeg\
56\arcmin\ 13\farcs 5 (J2000), close to the centroid from the RDCS
survey.  This centroid is south of the main peak in the X-ray
emission, at 08$^{\rm h}$ 48$^{\rm m}$ 47\fs 91 and $\delta$ =
44\arcdeg\ 56\arcmin\ 18\farcs 7 (J2000).  This shift in the centroid
from the main peak could be a result of flux from the peak to the
south-east.

Before we determined properties from the X-ray spectra, we needed an
estimate of the background spectrum.  We extracted from the merged
event list a nearby background region, which we cleaned of all point
sources.  We fit a combination of a broken power law model with a
Gaussian to represent the 2.1 keV Au emission line.  This is the same
model used in \citet{stanford00}, which examined two other clusters in
the same Chandra pointing.  We fit the model to the data using the
modified version of the statistic in \citet{cash79} recommended by
Castor in the XSPEC manual \citep{arnaud_xspec96}.  This statistic
does not require binning the data as it assumes the errors are
distributed in a Poisson manner, instead of as a Gaussian as expected
by a $\chi^{2}$ distribution.  The actual fit was performed over the
energy range 0.5--6.0 keV and the best fitting parameters agree well
with those found in \citet{stanford00}.

To estimate temperatures for the two different groups, we used two
30\arcsec\ radius apertures.  One was centered on the northern X-ray
peak while the other was centered on the secondary peak to the
southeast.  The apertures are plotted in Figure \ref{cl2_xray_cont}.
We excluded the region of the southern aperture that would have
overlapped with the northern aperture.  We fit each aperture
separately with the model of \citet{raymond77}, using the background
model described above after it was rescaled for the smaller area in
the two apertures.  For galactic absorption, we used a value of $2.6
\times 10^{20}$ cm$^{-2}$ from \citet{dickey90}.  We fit our objects
using the same energy range and statistic we used for fitting the
background region.

\begin{figure}
\begin{center}
\includegraphics[height=6.0in]{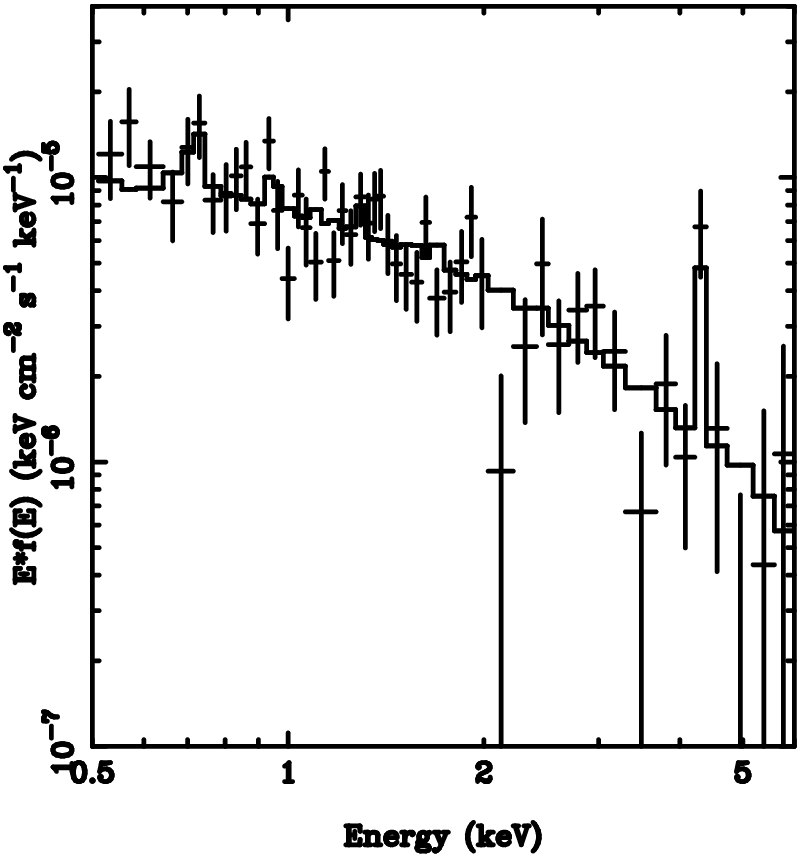}
\end{center}
\caption[cl2.fig3_pps.ps]{
The best fitting spectrum for the northern aperture plotted with the
binned events.  The events have been binned into groups of 20.
\label{north_sp}}
\end{figure}

\begin{figure}
\begin{center}
\includegraphics[height=6.0in]{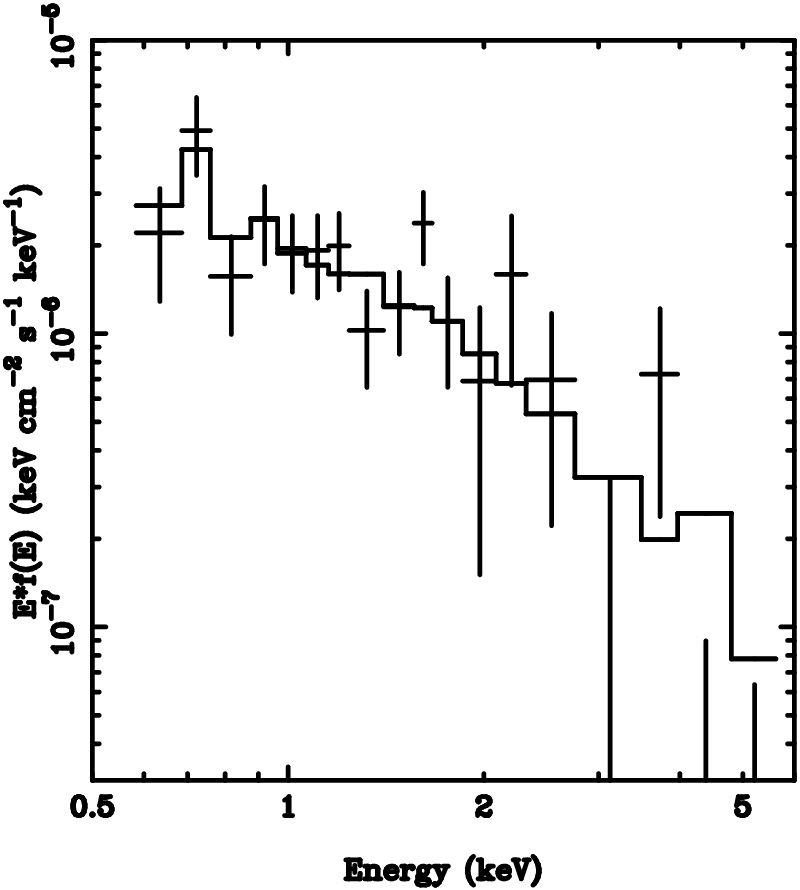}
\end{center}
\caption[cl2.fig4_pps.ps]{ Same as in Fig. 3, but the southern
aperture.  The much lower temperature of this model causes the 6.7 keV
line to disappear.
\label{south_sp}}
\end{figure}

When fitting the spectra, we allowed the abundance and the redshift to
be free parameters in addition to the normalization and the
temperature.  The northern aperture has a temperature of
$3.6^{+0.4}_{-0.4}$ keV, an abundance of $0.44^{+0.22}_{-0.18}$, and a
redshift of $z=0.572^{+0.013}_{-0.013}$.  We found for the southern
aperture a temperature of $2.3^{+0.5}_{-0.4}$ keV, an abundance of
$0.57^{+1.00}_{-0.39}$ and a redshift $z=0.569^{+0.044}_{-0.035}$.
For the northern aperture, the best fitting redshift was quite close
to the biweight redshift for the higher redshift grouping in \S 3.1
while $\simeq$ 2 $\sigma$ away from the lower redshift grouping.  The
abundances were typical of lower redshift clusters, though with very
larger errors on the southern aperture abundance measurement.  The
flux for the northern aperture is $2.9 \times 10^{-14}$ erg s$^{-1}$
cm$^{-2}$ (0.5--6.0 keV observed) while the flux for the southern
aperture is $0.6 \times 10^{-14}$ erg s$^{-1}$ cm$^{-2}$ (0.5--6.0 keV
observed). For the southern aperture, there are far fewer events
compared with the north, so the error estimates are larger.  These
results are summarized in Table \ref{xray_results} and we plot the
best fitting spectra in Figures \ref{north_sp} and \ref{south_sp}.

%

We fit a $\beta$ model, centered on the peak of the X-ray emission, to
the events from \CL\ from 0.5 -- 6.0 keV.  We excluded an aperture of
20\arcsec\ centered on the southern peak and smaller apertures around
nearby point sources.  We then binned the data into 40 circular
2\farcs 5 bins centered on the northern peak (not the centroid of the
X-ray emission).  Our largest aperture is at 100\arcsec\ so we could
simultaneously fit the model and the background.  We found the best
fitting model to have $\beta = 0.65 \pm 0.09$ and $r_{core} = 15\farcs
3 \pm 2\farcs 9$.  At a redshift of $z=0.570$, the core radius
corresponds to $r_{core} =108 \pm 21$ kpc.  These values of $\beta$
and $r_{core}$ are typical for clusters of galaxies \citep{jones99}.
We used these values, along with the spectral model above, to compute
a bolometric luminosity for the northern aperture.  We estimate this
to be $1.0 \pm 0.3 \times 10^{44}$ erg s$^{-1}$ for a cluster at
$z=0.570$.  The error quoted includes both the errors in the profile
fit as well as the errors in fit to the X-ray spectrum.  The
bolometric luminosity and measured temperature agree with expected
luminosity-temperature relation, as shown in \citet{borgani00}.

The temperatures of the northern and southern apertures differ at the
90\% confidence limit.  In other words, neither temperature falls
within the 90\% confidence limit of the other temperature, despite the
large errors for the southern aperture.  This suggests we could be
seeing X-ray emission from the two different groups.  Although the two
temperatures are not statistically inconsistent at the 3$\sigma$
level, it is tantalizing to have the two temperatures differ at the
90\% confidence limit, since this implies we are resolving the X-ray
emission from the two different systems nearly along the line of sight.

\subsection{Gravitational Arc}

In the HST imaging data, we have found a number of arc candidates. One
of these candidates is confirmed with a spectroscopic redshift.  We
find the arc at $\alpha$ = 08$^{\rm h}$ 48$^{\rm m}$ 48\fs 76 $\delta$
= 44\arcdeg\ 55\arcmin\ 48\farcs 61 (J2000) with $z=3.356$.  In Figure
\ref{cl2_xray_cont}, we mark the confirmed arc with a box.

\begin{figure}
\begin{center}
\includegraphics[height=6.0in]{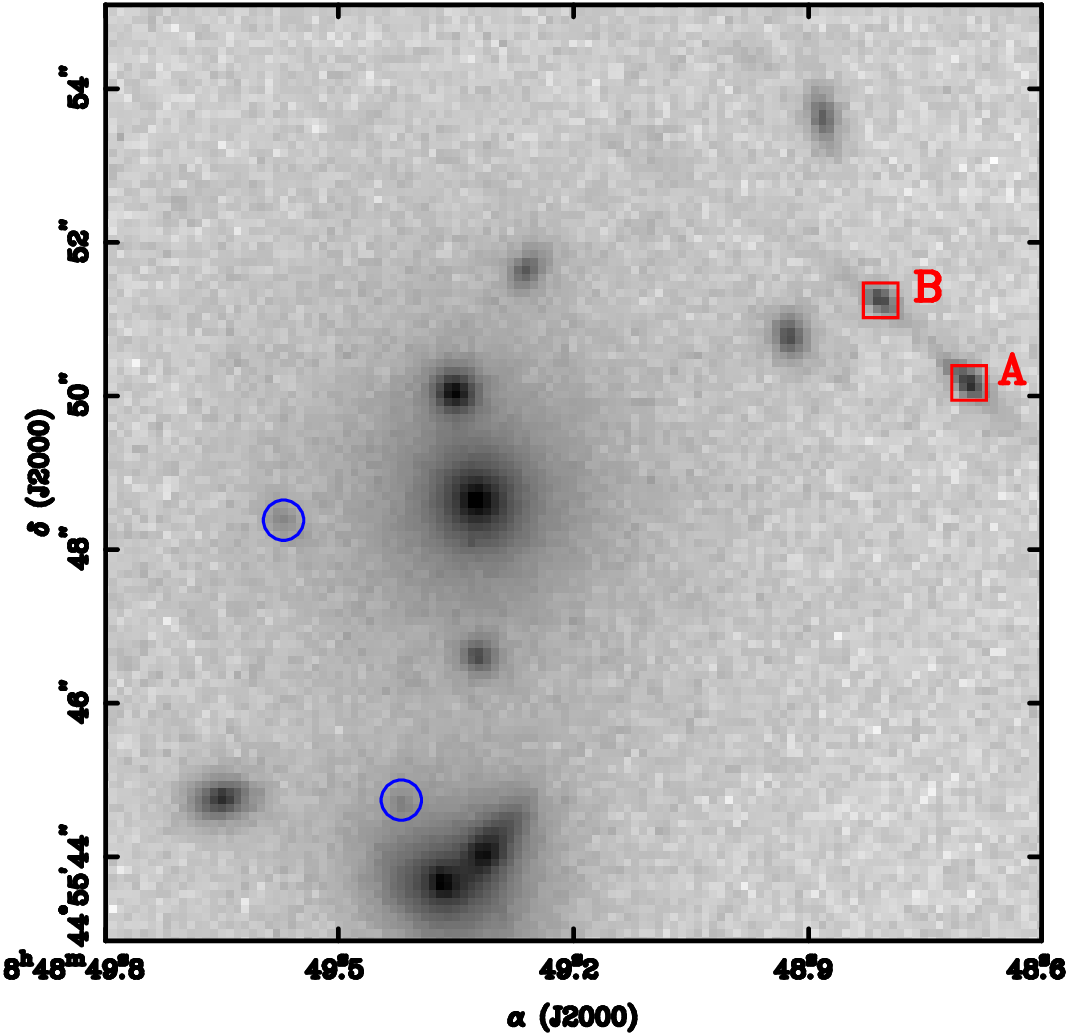}
\end{center}
\caption[cl2.fig5_pps.ps]{ Our HST image of the luminous elliptical
next to the $z=3.356$ gravitational arc. The arc has two bright
components, marked with squares A and B, both of which have identical
spectra as shown in Figure 6.  Two potential counter images are marked
with circles.  We display this image with a logarithmic scaling.
\label{bcg2}}
\end{figure}

The gravitational arc lies 6\farcs 1 from the $z=0.543$ elliptical
galaxy in the south, near the center of the southern X-ray peak.  We
show the arc and the elliptical galaxy in Figure \ref{bcg2}.  Based on
the northeast-southwest orientation of this arc, it is probably not
being lensed by the X-ray cluster at all, but rather by the elliptical
galaxy or a group of galaxies centered on the elliptical (see
Figure \ref{bcg2}).  

Also apparent in Figure \ref{bcg2} is that the arc has two components.
The LRIS spectra of the arc components extend from 5340 to 7895\AA\ in
the observed frame, and are plotted in Figure \ref{arc}. Both
components show a weak continuum and several strong emission lines: a
doublet around 6470\AA, a triplet near 6750\AA, a weak singlet at
7150\AA\ and another doublet around 7250\AA . There are no other
convincingly detected lines in the observed wavelength range. By
virtue of wavelength matching, these are readily identified, for a
redshift of $z=3.356$, as \ion{N}{4}] 1483\AA, 1486\AA; \ion{C}{4}
1548\AA, 1550\AA; \ion{He}{2} 1640\AA; and \ion{O}{3}] 1661\AA,
1666\AA.  The triple structure of the \ion{C}{4} resonance line then
suggests that there is some self absorption present. The lines are all
narrow with an observed FWHM = 180 km s$^{-1}$ which, when corrected
for instrumental resolution, implies an intrinsic width of $\sim$140
km s$^{-1}$. With the exception of \ion{O}{3}], the lines do,
however, show distinct broad bases.

While these emission lines are seen in objects ranging from quasars
and radio galaxies to planetary nebulae and symbiotic stars, their
relative ratios are, as far as we can discover from searching the
relevant literature and data archives, unique. In particular, the
\ion{O}{3}] doublet is relatively very strong while the
\ion{N}{3}] triplet at 1751\AA\ is only marginally detected. At this
redshift the Lyman $\alpha$\ line lies just outside the wavelength
covered by the spectra but we can confirm that the \ion{N}{5} 1238,
1242 resonance doublet is much weaker than \ion{N}{4}]. We confirmed
the wavelength of Lyman $\alpha$\ with a spectrum obtained using the
150 line mm$^{-1}$ grating as part of the Keck observations for
SPICES.

These spectra will be discussed in detail in a later paper, but we can
draw some preliminary conclusions about the nature of the source of
line emission from a qualitative examination of the spectrum. If the
gas is photoionized, the relative weakness of \ion{He}{2} and
\ion{N}{5} suggests that the ionizing spectrum is not as hard as a
typical AGN.  The absence of \ion{S}{4} and \ion{N}{3}, however,
suggest a high ionization parameter (U = photon/matter density). The
alternative hypothesis of shock excitation, both with and without
precursor ionizing photons \citep{dopita1996}, appears unable to
produce \ion{O}{3}] stronger than \ion{C}{4}. We conclude,
therefore, that the object being lensed is photoionized by a source
which is softer than a typical AGN but hotter than a black body of
$\sim 60,000$K, likely between 80,000 and 100,000 K. Establishing the
permissible range of element abundances must await detailed analysis
of all the spectral data, though the best fitting value is
$\simeq$0.05 ${\rm Z_{\sun}}$ with U$\simeq 0.1$.

The two components of the arc and the extended morphology imply that
the source object lies along a fold caustic, similar to an arc found
by \citet{kneib93} in Abell 370.  We have unsuccessfully searched for
possible counter images in the ground based I and K band data.  In
Figure \ref{bcg2}, we mark two possible counter images in the HST
image that would be too faint to appear in the ground based data.
Using the potential counter image to the East of the elliptical and
the assumption of an isothermal sphere, we derive a velocity
dispersion of 480 km s$^{-1}$, quite close to the group velocity
dispersion. As this model ignores the ellipticity of the lens, which
is necessary to produce the shape of the arc, the velocity dispersion
should be an over-estimate.

\section{Mass Estimates and Results}

We have good evidence that \CL\ consists of an X-ray emitting cluster
of galaxies at $z=0.570$, while there is an X-ray faint group or
low-mass cluster in front of \CL\ at $z=0.543$.  The centroid of all
the X-ray emission lies near four galaxies that are part of the
grouping at $z=0.570$, with the peak in the X-ray emission lying on
top of one of the bright, early-type galaxies in the $z=0.570$ system.
In addition, the best fitting redshift for the X-ray spectrum from the
northern component, $z=0.572^{+0.013}_{-0.013}$, is close to the
redshift of the $z=0.570$ grouping of galaxies.  These two pieces of
evidence point to \CL\ as the $z=0.570$ system.

The velocity dispersion-temperature relation of \citet{lubin93}
predicts a velocity dispersion of $710 \pm 110\ {\rm km\ s^{-1}}$ (the
errors are from the scatter in relation, not based on the temperature
errors) for a cluster with X-ray gas at kT = 3.6 keV.  We find that
the temperature of the gas near the centroid agrees well with the
measured velocity dispersion, or, in other words, that the dark matter
is in equilibrium with the X-ray emitting gas.  Using the relation
from \citet{borgani99}, which assumes a singular isothermal sphere, we
estimate the mass $3.0\ \pm\ 0.6 \times\ 10^{14}$ M$_{\sun}$
($z=0.570$ group) based on the velocity dispersion.  If we use the
X-ray temperature, we find a mass for the northern aperture of
$3.6^{+0.6}_{-0.6} \times 10^{14}$ M$_{\sun}$ using the
mass-temperature relation of \citet{Eke96}.  If we compare these mass
estimates with the low redshift mass-luminosity relation for X-ray
clusters \citep{wu99,reiprich00}, we find our bolometric luminosity in
good agreement with the values expected.  This confirms the lack of
strong evolution seen in the luminosity-mass relation for other RDCS
clusters in \citet{squires98}.

If we apply the same relations to the $z=0.543$ system, we find
results with far less agreement.  Using the same relation from
\citet{lubin93} and the velocity dispersion of the system of galaxies
at $z=0.543$, we would expect kT = $1.6 \pm 0.4$ keV for the southern
region, which is less than the X-ray temperature measurement by 1.8
standard deviations.  If we use the \citet{borgani99} relation we find
a mass of $8.0\ \pm 1.1\ \times\ 10^{13}$ M$_{\sun}$ while the
\citet{Eke96} mass-temperature relation predicts $1.9^{+0.6}_{-0.5}
\times 10^{14}$ M$_{\sun}$, more than two standard deviations from the
mass estimated from the velocity dispersion.  Whichever of the two
mass estimates we use, we find a small binding mass for the $z=0.543$
group.  Suffice it to say, we know that this lower mass system is a
real group of galaxies and not a projection because of the
gravitational lens.  Detailed modeling of the gravitational lens
should yield a better estimate for the mass of the lower redshift
system.

We find it likely that the southern aperture is contaminated from
X-ray emission by the northern X-ray cluster, so there would be a
systematic error in the temperature estimate that could explain the
discrepancies above.  One reason why we suspect contamination is the
high redshift found for the southern aperture, a redshift closer to
the $z=0.570$ system than to the $z=0.543$ system.  Secondly, if we
use the results from the $\beta$ model fit, we compute that we expect
87 events to land within the southern aperture which is almost half of
the total number of events.

What is peculiar is that there are a fairly large number of galaxies
in the lower redshift group.  We measured 17 redshifts (though five
are emission line galaxies), more than the 11 in the $z=0.570$
cluster.  Including the 150 line mm$-1$ grating redshifts further
bears this out.  We find 32 galaxies in the $z=0.543$ group and 25 in
the $z=0.570$ cluster.  As we do not have a complete sample from the
SPICES survey, we can not make any firm statistical statements.
However, these numbers are entirely consistent with a roughly equal
distribution of galaxies in the two objects.  So it appears there are
a fair number of galaxies associated with a group that has a small
binding mass and little or no X-ray emission.  With the completed
SPICES data, we will be able to study the luminosity function of these
two different systems and thus confirm the apparent high richness of
this low luminosity system.

One question that we need to address is what sort of gravitational
potential is causing the arc?  There are a number of different models
that need to be explored to fully characterize this system, which will
be done in a later paper.  The large separation of the lens and the
elliptical, however, makes it unlikely that just the galaxy itself is
lensing the source.

\section{Summary}

\CL\ appears to be a relatively complicated system, when viewed from
our line of sight.  We find it most likely that the majority, though
not all, of the X-ray emission comes from a cluster of galaxies at
$z=0.570$, which is \CL\ proper.  The peak of the X-ray emission is
close to the brightest cluster member.  The best fitting redshift for
the X-ray spectrum is $z=0.572$, less than one standard deviation from
the peak in the galaxy redshift distribution at $z=0.570$ but more
than two standard deviations from the $z=0.543$ grouping of galaxies.

The cluster has an X-ray temperature of $3.6^{+0.4}_{-0.4}$ keV and a
velocity dispersion of 670 $\pm$ 50 ${\rm km\ s^{-1}}$.  These two
values result in a $\beta_{spec}$, the ratio of the kinetic energy
per unit mass in the galaxies to that in the X-ray gas, very close to
one.  This means that the dark matter and the X-ray gas are very close
to equilibrium, for the central parts of the cluster at least, which
implies this cluster is mostly relaxed.

In front of \CL, at $z=0.543$, there is a group or low mass cluster of
galaxies with a velocity dispersion of 430 $\pm$ 20 ${\rm km\
s^{-1}}$.  Though it appears there are a fairly large number of
galaxies that are members of this system, it has a much smaller X-ray
luminosity than \CL.  We found, in the aperture centered on the
southern peak, an X-ray flux \slantfrac{1}{5}\ of that in \CL, though
after correcting for contamination from \CL, a flux of only
\slantfrac{1}{10}\ of that in \CL\ is more likely.  This
group at $z=0.543$ contains a gravitational lens, a useful probe of
the potential of this slightly lower redshift group.

\CL\ illustrates why detailed follow up on clusters of galaxies is
important.  A large aperture, covering both the cluster and the
foreground group, produces a lower temperature.  For example, an
aperture of 50\arcsec\ centered on the X-ray centroid produces a
temperature of kT = $2.9^{+0.2}_{-0.2}$ keV, almost two standard
deviations lower than the temperature we measured above.  Without the
velocity information showing that there are two groups along the line
of sight, we would be more inclined to accept this lower value.
Without the high resolution X-ray data, we would not know which of the
two groups actually was the source of the original X-ray emission.
The combination of optical spectroscopy, which allowed us to resolve
the two different systems along the line of sight, and the high
spatial and spectral resolution X-ray data, gave us the necessary data
to disentangle this cluster.

We plan to continue observing RDCS clusters in this multi-wavelength
fashion.  This will allow us to not only understand the distribution
of masses of clusters of galaxies, but also study the galaxy
properties when we combine high resolution HST imaging and Keck
spectroscopic data.  With mass estimates from a variety of techniques,
we will be able to construct a clean sample of clusters of galaxies
for comparisons with theoretical expectations.

BH would like to thank Michael Gregg and Patrick Wojdowski for useful
discussions.  We thank Scott Wolk for assistance with planning our
Chandra observation.  We thank Montse Villar-Martin for discussions
about the arc emission line spectrum.  We would also like to thank the
referee for helping us make this a better paper.  Support for SAS came
from NASA/LTSA grant NAG5-8430 and for BH from NASA/Chandra GO0-1082A.
Both BH and SAS are supported by the Institute for Geophysics and
Planetary Physics (operated under the auspices of the US Department of
Energy by the University of California Lawrence Livermore National
Laboratory under contract W-7405-Eng-48). Support for GKS in this work
was provided by NASA through Hubble Fellowship Grant
No. HF-01114.01-98A from the Space Telescope Science Institute, which
is operated by the Association of Universities for Research in
Astronomy, Incorporated, under NASA Contract NAS5-26555.  RAEF is
affiliated to the Astrophysics Division of the Space Science
Department, European Space Agency.  Some of the data presented herein
were obtained at the W.M. Keck Observatory, which is operated as a
scientific partnership among the California Institute of Technology,
the University of California and the National Aeronautics and Space
Administration.  The Observatory was made possible by the generous
financial support of the W.M. Keck Foundation.

\begin{figure}
\caption[cl2.fig1_pps.ps]{ 
The X-ray contours overlaid on the WFPC2 F702W data with north up and
east to the left.  The X-ray contours are from the 0.5--6.0 keV events
smoothed with a 2\arcsec\ FWHM Gaussian.  The gravitational arc is
marked with a small box.  The apertures we used for spectral
extraction are marked with circles.  All four of the bright galaxies
to the east of the peak in the X-ray emission are part of the $z=0.570$
group.  The elliptical galaxy near the fainter X-ray peak, to the
east of the green box, is at a redshift of $z=0.543$.  The contour
levels are 0.9, 1.4, 1.9, 2.7, 3.4 and 4.4 $\times 10^{-17}$ erg
s$^{-1}$ cm$^{-2}$ arcsec$^{-2}$.
\label{cl2_xray_cont}}
\end{figure}

\begin{figure}
\caption[cl2.fig6_pps.ps]{ 
The rest frame spectra for the two arc components at a redshift of
$z=3.356$.  These are not flux calibrated and are plotted as raw CCD
counts. For comparison, we plot $F_{\lambda}$\ spectra covering the
same wavelength range of an LMC planetary nebula, LMC SMP-47, observed
with the HST Faint Object Spectrograph, \citep{vassiliadis1996} and of
an average $z\sim 2.5$\ radio galaxy \citep{vernet2001}. Note the
strong intercombination lines of nitrogen and oxygen in the arc. Noise
residuals from the bright night sky lines at observed wavelengths of
5577, 5893, 6300 and 6363 \AA\ are marked.
\label{arc}}
\end{figure}

\tablenum{1}
\begin{deluxetable}{lcccccc}
\tablecaption{Results of Fitting the X-ray Spectra\label{xray_results}}
\tablehead{
\colhead{Aper.} & \colhead{$\alpha$} & \colhead{$\delta$}
& \colhead{Net Counts} & \colhead{kT} & \colhead{Abund.} & \colhead{z} \\ 
\colhead{} & \colhead{(J2000)} & \colhead{(J2000)} & \colhead{(0.5-6.0
keV)}
& \colhead{(keV)} & \colhead{} & \colhead{} \\ 
}
\startdata
North & 08$^{\rm h}$ 48$^{\rm m}$ 48\fs 04 &
 +44\arcdeg\ 56\arcmin\ 20\farcs 42 & 851.6
& $3.6^{+0.4}_{-0.4}$ & $0.44^{+0.22}_{-0.18}$ &
$0.572^{+0.013}_{-0.013}$ \\[3pt]
South & 08$^{\rm h}$ 48$^{\rm m}$ 49\fs 25 &
 +44\arcdeg\ 55\arcmin\ 48\farcs 46 & 185.1
& $2.3^{+0.5}_{-0.4}$ & $0.57^{+1.00}_{-0.39}$ &
 $0.569^{+0.044}_{-0.035}$ \\
\enddata
\end{deluxetable}

\end{document}